\documentstyle[amssymb,preprint,eqsecnum,aps,prb]{revtex}

\begin{document}
\title{Magnetic- field -induced structural changes in the electron doped manganites
Ca$_{x}$Sm$_{1-x}$MnO$_{3}$ (x = 0.8, 0.85) }
\author{R. Mahendiran$^{1,2}$, P. A. Algarabel$^1$, L. Morellon$^1,$ C. Marquina$^1$%
, M. R. Ibarra$^1$,}
\author{A. Maignan$^2$, C. Martin$^2$, \ M. Hervieu$^2$, B. Raveau$^2$, and C. Ritter%
$^3$}
\address{$^1$Departamento de F\'{i}sica de la Materia Condensada-ICMA,\\
Universidad de Zaragoza-CSIC, 50009, Zaragoza, Spain\\
$^2$Laboratoire CRISMAT, ISMRA, Universit\'{e} de Caen, 6 Boulevard du\\
Mar\'{e}chal Juin, 14050 Caen-Cedex, France\\
$^{3}$Institut Laue-Langevin, Boite Postale 156, 38042, Grenoble Cedex 9,\\
France}
\date{\today}
\maketitle

\begin{abstract}
We studied the correlation between magnetic, electrical, structural, and
magnetostriction properties of the electron doped manganites Ca$_{x}$Sm$%
_{1-x}$MnO$_{3}$ (x = 0.85, 0.8). The paramagnetic to antiferromagnetic
transition in both the compounds while cooling is accompanied by an abrupt
increase of the spontaneous volume thermal expansion ($\Delta $V/V = 0.07 \%
for x = 0.85 and 0.25 \% for x = 0.2). The x = 0.15 exhibits multiple phase
separation at 5 K: G-type, and C-type antiferromagnetic phases in
orthorhombic ({\it Pnma}) and monoclinic ({\it P2}$_{1}${\it /m}) structures
respectively. Magnetic study on x = 0.85 also suggest ferromagnetic regions
possibly in $Pnma$ structure coexist with the antiferromagnetic phases. \
The magnetization (M = 1.2 $\mu _{B}$) of x = 0.85 does not reach the value
expected for the complete alignment of \ Mn spins even at 12 T and at 12 K.
Metamagnetic transitions (C-type to Ferromagnetic) in both compounds are
accompanied by contraction of volumes under high magnetic fields. We suggest
\ that a high magnetic field induces {\it P2}$_{1}${\it /m} (high volume) to 
{\it Pnma} (low volume) structural transition. This is also supported by the
neutron diffraction study.
\end{abstract}


\section{INTRODUCTION}

In spite of extensive investigations on the hole doped, Mn$^{3+}$: t$_{2g}^3$%
e$_g^1$ rich manganites\cite{Coey} there are scarcely few reports on the
electron doped, Mn$^{4+}$:t$_{2g}^3$e$_g^0$ rich compounds. Recently
electron-doped CaMnO$_3$ compounds regained experimental$^{2-9}$ and
theoretical interests$^{10}$ motivated by the following facts:.1. The
compounds x 
\mbox{$>$}%
0.5 in the series RE$_{1-x}^{3+}$Ca$_x^{2+}$MnO$_3$ (RE = La, Nd, Pr, etc.,)
are antiferromagnetic insulators and the antiferromagnetic state is
insensitive to the external magnetic field of $\mu _0$H = 5-7 T normally
employed in many laboratories\cite{Rao}. However, \ magnetoresistance of
about 82 \% was first reported by Mahendiran et al\cite{Mahi96} in La$_{0.1}$%
Ca$_{0.9}$MnO$_3$ (Mn$^{4+}$ = 85 \%) around the Neel temperature (T$_N$).
2. Investigations by Chiba et al\cite{Chiba96} on Bi$_{1-x}$Ca$_x$MnO$_3$
and by Maignan et al\cite{Martin97} on RE$_{1-x}$Ca$_x$MnO$_3$ (RE = Nd, Sm
Gd, etc.,) revealed that magnetoresistance exists over a wide range (0.96 $%
\leq $ x $\leq $ 0.8) and the maximum magnetoresistance was again obtained
for Mn$^{4+}$ $\approx $ 85 \%. In contrast to the hole-doped (rich in Mn$%
^{3+}$:t$_{2g}^3$e$_g^1$) compounds whose resistivity ($\rho $) becomes
insensitive to a fixed external magnetic field as size of the rare earth ion
decreases\cite{Coey,Rao}, a large magnetoresistance independent of the rare
earth ion is observed in the electron doped manganites\cite
{Mahi96,Chiba96,Martin97}. 3. Magnetization of the electron doped compounds
for 0.85 $<$ x $<$ 1 surprisingly increases below T$_N$\cite
{Mahi96,Chiba96,Martin97,Neum} whose origin is attributed to the formation
of ferromagnetic clusters and their growth with x in the G-type
antiferromagnetic background although canted antiferromagnetic structure for
certain x value cannot be completely ruled out\cite
{Mahi00,Neum,Santhosh,Savotsa}. Interestingly $\rho $ also exhibits a
maximum around T$_N$ in compounds (x = 0.97-0.88) which have higher
magnetization \cite{Martin97}. 4. Electron-doped manganites serve as a
laboratory to study the dynamics of excess electrons (e$_g$ electrons of Mn$%
^{3+}$:t$_{2g}^3$e$_g^1$) in an antiferromagnetic matrix with
compositionally tunable Jahn-Teller interactions and model to verify
electronic phase separation proposed theoretically\cite{Moreo}. When
intersite Jahn-Teller interactions becomes dominant the G-type simple two
sublattice antiferromagnetic order in Sm$_{1-x}$Ca$_x$MnO$_3$ (x = 1)
changes to\ the C-type antiferromagnet (ferromagnetic linear chains with
inter chain antiferromagnetic coupling) for x = 0.87-0.8 without charge
ordering\cite{Martin99} and with charge ordering for 0.8 
\mbox{$<$}%
x 
\mbox{$<$}%
0.6\cite{Hervieu}. The composition x = 0.5 is generally a CE-type
antiferromagnet (zig-zag ferromagnetic chains with inter chain
antiferromagnetic coupling) with charge and cooperative Jahn-Teller ordering
occurring at T 
\mbox{$>$}%
T$_N$. Most of the existing studies have attributed the large
magnetoresistance in manganites to the magnetic field induced ferromagnetic
alignment of the t$_{2g}^3$ core spins either from the antiferromagnetic or
the paramagnetic state although few investigations\cite{Ibarra1,Booth}
suggest that magnetically tunable lattice distortion could play an important
role. However, the sign of lattice deformations either in zero or non zero
magnetic field \ is not predicted theoretically so far and the existing
experimental results are mostly on the \ hole doped (x 
\mbox{$<$}%
0.5) or x = 0.5 compounds\cite{Ibarra1}. The purpose of this paper is to
investigate the nature of \ the lattice distortion in absence and in
presence of a external magnetic field in the electron doped Sm$_{1-x}$Ca$_x$%
MnO$_3$.series (x = 0.85, 0.8) which undergo simultaneous magnetic
(paramagnetic to a C-type antiferromagnetic) and structural transitions
(orthorhombic to monoclinic) with lowering temperature\cite{Hervieu}.

\section{EXPERMENT}

We measured magnetization of polycrystalline samples of Ca$_x$Sm$_{1-x}$MnO$%
_3$ (x = 0.85, 0.8) using a commercial vibrating sample magnetometer up to
12 T and zero field four probe resistivity was also measured in the
temperature range 5K 
\mbox{$<$}%
T 
\mbox{$<$}%
300 K. The structural characterization of our samples was already reported 
\cite{Martin99,Hervieu}. Linear thermal expansion ($\Delta $L/L) in zero
magnetic field was measured for 20 K 
\mbox{$<$}%
T 
\mbox{$<$}%
300 K using the strain gauge technique. The volume thermal expansion was
calculated using $\Delta $V/V = 3$\Delta $L/L. Magnetostriction by the
strain gauge method was measured in a pulsed magnetic field up to 14 T with
the magnetic field parallel and perpendicular to the measuring direction.
The volume magnetostriction of randomly oriented polycrystallites was
calculated through $\Delta $V/V = $\lambda _{par}$+2$\lambda _{per}$ where $%
\lambda _{par}$ and $\lambda _{per}$ are parallel and perpendicular
magnetostrictions. The anisotropic magnetostriction was found to be
negligible and not presented in this paper. Neutron powder diffraction
spectra at T = 100 K in 0 T and 6 T on Ca$_{0.85}$Sm$_{0.15}$MnO$_3$ with Sm$%
^{154}$ isotope were carried out in the high resoultion spectrometer (D2B
instruement ) at Institue Laue-Langvein (ILL),Grenoble using the wavelength
of 1.594 Angstrom.

\section{RESULTS\AA}

Fig. 1(a) show temperature dependence of the inverse magnetic susceptibility
(H/M.) for x = 0.85 and 0.8 while cooling from 300 K. The arrows mark the
onset of C-type antiferromagnetic ordering (T$_{NC}$ = 112 K for x = 0.85
and 150 K for x = 0.8). The paramagnetic to a C-type antiferromagnetic
transitions are accompanied by orthorhombic (${\it Pnma}$ space group) to
monoclinic ({\it P2}$_{{\it 1}}${\it /m}) structural transitions\cite
{Martin99,Hervieu}. The magnetic transition in these compounds are
hysteretic as shown already\cite{Mahi00}. Earlier neutron diffraction study
on x = 0.85\cite{Martin99} showed that in addition to the C-type ordering in
the monoclinic phase, another antiferromagnetic ordering of G-type also sets
in orthorhombic phase around T$_G$ = 118 K and its phase fraction increases
from 5\% at 100 K to 9 \% at 10 K. Due to its smaller phase fraction and
closeness of T$_G$ to T$_C$, its signature is not very clear in the magnetic
measurements. The Curie-Weiss fit (dark lines) for T 
\mbox{$>$}%
T$_{NC}$ gives Curie-Weiss temperatures $\theta _p$ =119 K ( x = 0.85) and
135 K (x = 0.8). The experimentally determined effective moments P$_{eff}$ =
4.386$\mu _B$ (x = 0.85), 4.238$\mu _B$ (x = 0.8) are slightly higher than
the theoretical values P$_{eff}$ = 4.162 $\mu _B$ for x = 0.85 and 4.215 $%
\mu _B$ for x = 0.8. The x = 0.8 composition is expected to undergo a
paramagnetic to a C-type antiferromagnetic transition below T$_{NC}$ = 150 K
without coexistence of phases at low temperature based on the measurements
on similar compounds\cite{Bao97,Santhosh}

\bigskip

Fig. 1(b) shows the zero field resistivity $\rho $(T) of x = 0.85 and x =0.8
while cooling from 300 K to 5 K and warming back to 300 K. We clearly see
hysteresis in resistivity around T$_{NC}$. The crystallographic transition
accompanying magnetic transition are marked. At 300 K, $\rho $(x = 0.8) $>$ $%
\rho $(x = 0.85). In contrast to the insulating behavior of $\rho $(T) in
the paramagnetic phase of hole-doped compounds\cite{Mahi96}, $\rho $(T) of
these two electron-doped compounds decreases as the temperature decreases
from 300 K and shows a broad minimum at T$_{min}$ = 155 K ( x = 0.85), 210 K
(x = 0.8) as shown in the inset and then jumps at T$_{NC}$ ( 
\mbox{$<$}%
T$_{min}$). In between T$_{N}$ and 5 K, $\rho $(T) increases by four orders
of magnitude in x = 0.85 and eight orders of magnitude in x = 0.8. The
resistivity minimum above T$_{NC}$ can be recaptured in non adiabatic
polaronic picture when the thermal and polaron activation energies are
equal. We were able to fit the data using the relation $\rho $(T) = $\rho
_{0}$T$^{1.5}$exp(E$_{a}$/k$_{B}$T) with E$_{a}$ = 0.017 eV and E$_{a}$ =
0.023 eV. With increasing content of the the rare earth ion, T$_{min}$
shifts to much higher temperatures\cite{Wool}

\smallskip

Fig. 1(c) shows the volume thermal expansion ($\Delta $V/V) with respect to
V value at 300 K. As T decreases from 300 K, the lattice shrinks in the
paramagnetic region and then shows a rapid increase at T$_{NC}$. In the
entire temperature range (300 K- 20 K), the over all change in $\Delta $V/V
is larger for x = 0.85 than x = 0.8. However, around T$_{NC}$ the volume of
x = 0.8 increases by 0.35 \% which is five times larger than in x = 0.85 ( $%
\Delta $V/V $\thickapprox $ 0.07 \% around T$_{NC}$). These differences are
intimately tied to the degree of localization of charges (for example,
resistivity of x = 0.8 at 5 K is 5 orders of magntidue higher than x = 0.8),
the strength of antiferromagnetic exchange interactions and structural
distortions. Here, it should be mentioned that the magnitude of \ $\Delta $%
V/V in x = 0.85 is very senstive to the stoichiometry\cite{Pedro}. All
previous studies on manganites undergoing transition either from a
paramagnetic insulator to a ferromagnetic metal or from a ferromagnetic
metal to a charge-ordered antiferromagnetic insulator showed that volume
contracts at the transition temperature \cite{Ibarra1,Ritter} in contrast to
the volume expansion behavior seen here.

\smallskip

Figs. 2(a)-(d) illustrate the influence of external magnetic fields on the
spin and lattice degrees of freedom. The field dependence of the
magnetization and the magnetostriction data were taken after zero field
cooling to a predetermined temperature. At 12 K (T 
\mbox{$<$}%
\mbox{$<$}%
T$_{NC}$) M(H) of x = 0.85 [Fig. 2(a)] shows a rapid increase at low fields
( $\mu _{0}$H 
\mbox{$<$}%
0.5 T), gradually increases up to 2.5 T and then shows a rapid, but a smooth
increase until the maximum field of $\mu _{0}$H = 12 T. The trend of the
high field data suggest that M(H) can increase even beyond $\mu _{0}$H = 12
T. Upon decreasing H, M(H) exhibits a huge hysteresis. The low field
behavior is caused by the presence of ferromagnetic regions in
antiferromagnetic phase as suggested previously\cite{Mahi00}. The rapid
increase of M(H) above 2.5 T is due to the field induced transition from the
C-type AF to ferromagnetic state. It should be noted that the maximum
magnetic moment at 12 K even at a field as high as 12 T is 1.45 $\mu _{B}$
which is much less than 3.1 $\mu _{B}$ expected for a fully ferromagnetic
alignment of Mn spins. This means only a part of the sample transforms into
ferromagnetic state and the rest remaining in antiferromagnetic state. We
will touch upon this aspect in the later part of the paper. As T increases
the width of hysteresis decreases and finally at 175 K and 225 K, M(H) is
linear in H as expected for a paramagnet. Fig. 2(b) shows M(H) of x = 0.8
very close but below T$_{N}$. First we note that the ferromagnetic like low
field behavior seen in x = 0.85 is absent in x = 0.8. Second, the
metamagnetic transition is rather sharp and the width of the hysteresis loop
is nearly same (= 1.5 T) for T = 140 K and 145 K. The magnetic moment even
at a field as high as 12 T is $\approx $ 1.2 $\mu _{B}$ (see 145 K data)
which is far below the value of \ 3.2 $\mu _{B}$ expected for a fully
ferromagnetic alignment of Mn$^{3+}$ and Mn$^{4+}$ spins. The critical field
H$_{C}$ for the metamagnetic transition, as determined from the inflection
point of the rapidly increasing part of M(H) in the field-up mode increases
rapidly from 8 T to 11 T as the temperature decreases by 5 K from 145 K.

\bigskip

The field-induced metamagnetic transition is also accompanied by a dramatic
change in volumes. The volume magnetostriction ($\omega $) isotherms in Fig.
2(b) and 2(d), for x = 0.8 and x = 0.85 respectively, show that the lattice
volume contracts upon transition from the antiferromagnetic to the
metamagnetic state. The reverse transition is hysteretic. The maximum volume
change at 13.7 T is -0.135 \% at 25 K for x = 0.85 and decreases to -0.085
\% at 100 K. As like the magnetization, the hysteresis in $\omega $ also
decreases with increasing temperature. At 120 K, $\omega $ is positive and
is 0.0185 \%. The volume change is negligible for x = 0.8 at 12 K. An
appreciable magnetostriction in x = 0.8 appears only in between 120 K and
150 K. The maximum value of magnetostriction ($\Delta $V/V)$_{max}$
increases with H, reaches a maximum at 135 K and then decreases again. Fig.
3 is complementary to Figs. 2(c) and (d) and it compares the values of the
volume magnetostriction at the maximum field $\mu _{0}$H = 13.7 T for x =
0.85 and 0.8.as a function of temperature. The ($\Delta $V/V)$_{max}$ is
zero far away from T$_{NC}$ and at very close but above T$_{NC}$ it has
smaller positive value and, a rapid increase occurs just below T$_{NC}$. The
peak in ($\Delta $V/V)$_{max}$ for x = 0.8 is due to the fact that the
maximum available field is insufficient to induce a complete
antiferromagnetic to ferromagnetic transition.

\section{Discussion}

Now we turn attention to the origin of the volume contraction under a
magnetic field. The paramagnetic to C-type AF transition in the electron
doped compounds under investigation is accompanied by orthorhombic ($Pnma$)
- monoclinic (P2$_{1}$/m) structural transition\cite{Martin99}. The C-type
antiferromagnetism is coupled to polarization of the d$_{z^{2}-r^{2}}$
orbitals of Mn$^{3+}$ sites along the c-axis\cite{Martin99}. The x = 0.85
composition exhibits multiple phase separation. It was found that about 9 \%
of the G-AF phase in ${\it Pnma}${\it \ }structure also coexists with the
C-phase (P2$_{1}$/m structure) at 10 K in zero field in x = 0.85.\ In
addition, a third phase which is ferromagnetic also appears to be present in
x = 0.85\cite{Mahi00}. But the volume fraction of the ferromagnetic phase is
extremely sensitive to sample preparation\cite{Pedro}. The ferromagnetic
phase is in ${\it Pnma}$ structure, if one extrapolates the trend seen in x
= 0.9 compound\cite{Savotsa}. When H is increased from 0 T, the magnetic
contribution due to domain wall displacement in ferromagnetic phase
dominates at low fields. As H increases further above 5 T at 12 K, a
metamagnetic transition takes place and a portion of the C- AF phase is
converted into a ferromagnetic phase. Since, the magnetic moment does not
saturate even at the highest field of 12 T, ferromagnetic phase and
antiferromagnetic phase coexists in between 4 T and 12 T at 12 K.\ Because
of the strong coupling between the C-type AF and d$_{z^{2}-r^{2}}$ orbitals,
the destruction of the C-type AF is accompanied by disordering of d$%
_{z^{2}-r^{2}}$ which \ drives $P2_{1}/m$ to $Pnma$ structural transition.
Since the {\it P2}$_{1}${\it /m} phase higher volume than the {\it $Pnma$}
phase, the field induced monoclinic to orthorhommbic transition leads to a
volume contraction as seen in the magnetostriction data (Fig. 2(c)). If this
is to true, we can expect\ to see an increase in the {\it $Pnma$} phase
fraction above at high fields. This is what indeed found in the neutron
diffraction study\cite{Pedro}. While small angle neutron scattering study
and elaborate powder neutron diffraction study in Ca$_{0.85}$Sm$_{0.15}$MnO$%
_{3}$ will be published by some of us elsewhere\cite{Pedro}, we show the
result most relevant to this work. Fig. 4 shows the schematic diagram of the
nuclear phases of x = 0.85 at 100 K in absence of a magnetic field and in
presence of H = 6 T. The $P2_{1}/m$ phase (C-AF) which is 91 \% at 0 T
reduces to 44 \% at 6 T and correspondingly the $Pnma$ phase (G-AF)
increases from 9 \% at 0 T to 56 \% at 6 T. A large ferromagnetic component
of = 1.7 $\mu _{B}$ is also found in the $Pnma$ phase\cite{Pedro}. We
believe that similar field induced structural transition also takes place in
x = 0.8 compound.

\bigskip

\section{SUMMARY}

In summary, we have shown that the paramagnetic to the C-type
antiferromagnetic transition in the electron doped manganites Ca$_x$Sm$%
_{1-x} $MnO$_3$ (x = 0.80, 0.85) is accompanied by an abrupt volume
expansion ($\Delta $V/V = 0.07 \% in x = 0.85, 0.35 \% in x = 0.8) and
orthorhombic (Pnma) to monoclinic (P21/m) structural transitions. The x =
0.85 compound is magnetically and structurally phase separated with two
types of antiferromagnets ,G-(Pnma) and C- (P21/m) and a minority
ferromagnetic phase (Pnma) coexisting together. Magnetization study shows
that metamagnetic transiton (C-AF to ferromagnetic) field increases from 5 T
in x = 0.85 to more than 12 T in x = 0.8 at 12 K. Isothermal
magnetostriction measurements shows high volume to low volume transition
under a magnetic field which has been suggested to structural transition ($%
P2_1/m$ to $Pnma$ ) under a field. The driving factor for this structural
transition is the strong coupling between the C-type antiferromagnetic order
and the d$_{3z^2-r^2}$ orbital order. The field induced structural
transition in x = 0.85 compound is supported by neutron diffraction study.

\section{Acknowledgments}

R. M thanks MENRT (France) and Ministerio de Ciencia Y Cultura (Spain) for
financial assistance and acknowledges discussions with Prof. T. V.
Ramakrishnan and Dr.Venkatesh Pai on electron doped manganites\newpage
\newpage

\begin{center}
{\bf Figure captions}
\end{center}

\begin{description}
\item[Fig. 1]  Temperature dependence of the inverse susceptibility (H/M)
measured under $\mu _0$H = 0.01 T in Ca$_x$Sm$_{1-x}$MnO$_3$ (x = 0.8,
0.85). G- and C-type antiferromagnetic phases coexist down to the lowest
temperature in x = 0.15. PM -paramagnetic phase. (b). Temperature dependence
of the zero field resistivity of x = 0.85 and x = 0.8 while cooling and
warming. The inset shows the resistivity minimum in the paramagnetic state
(c). Temperature dependence of the spontaneous volume thermal expansion of x
= 0.8 and x = 0.85. The structural transitions are also marked.

\item[Fig. 2]  Magnetization isotherms of (a) x = 0.85 and (b) x = 0.8.
Volume magnetostriction isotherms of \ (c) x = 0.15 and (d) x = 0.8.

\item[Fig. 3]  Temperature dependence of the volume magnetostriction ($%
\Delta $V/V) of x = 0.85 and 0.8 at the the maximum field $\mu _0$H = 13.7 T.

\item[Fig. 4]  A schematic diagram of the volume phase fractions of \
nuclear phases in 0 T and 6 T at 100 K in Ca$_{0.85}$Sm$_{0.15}$MnO$_3$.

\item  \newpage 
\end{description}

\end{document}